\documentclass[twocolumn,aps]{revtex4}
\usepackage{epsf}
\begin{document}
\title{Leadership Statistics in Random Structures}
\author{E.~Ben-Naim}
\email{ebn@lanl.gov} \affiliation{Theoretical Division and Center
for Nonlinear Studies, Los Alamos National Laboratory,  Los
Alamos, New Mexico, 87545}
\author{P.~L.~Krapivsky}
\email{paulk@bu.edu} \affiliation{Center for Polymer Studies and
Department of Physics, Boston University, Boston, MA, 02215}
\begin{abstract}
The largest component (``the leader'') in evolving random structures
often exhibits universal statistical properties.  This phenomenon is
demonstrated analytically for two ubiquitous structures: random trees
and random graphs. In both cases, lead changes are rare as the average
number of lead changes increases quadratically with logarithm of the
system size. As a function of time, the number of lead changes is
self-similar. Additionally, the probability that no lead change ever
occurs decays exponentially with the average number of lead changes.
\end{abstract}
\maketitle 

Extreme statistics are important in science, engineering, and society
as they dictate catastrophic events, systems robustness, financial
indices, etc. The theory of extreme statistics provides a powerful
analysis framework and prediction tool \cite{res,jg}. However, it is
limited to ensembles of {\em independent} random variables.  Even
though most practical applications involve correlated random
variables, such cases remain largely unexplored \cite{bm,bkm,mk}.

We investigate extremal characteristics of two basic random
structures: random trees and random graphs. Random trees appear in
data storage algorithms in computer science \cite{hmm,dek,ws} and in
physical processes such as collisions in gases \cite{vvd}.  Random
graphs \cite{bb,jlr} have numerous applications to theoretical
computer science \cite{dek,ws}, social networks \cite{sp,gn}, and
physical processes such as polymerization \cite{pjf}.

We focus on the largest component, the leader, and ask: What is the
size of the leader?  How does the number of lead changes depend on
time and system size?  What is the probability that the leader never
changes? Similar questions were investigated in growing networks
\cite{mda,kr}, and related leadership statistics were studied in
random graphs by Erd\H os and \L uczak \cite{tl}.

Random trees and random graphs are special cases of aggregation
processes and hence, we analyze them using the rate equation approach
\cite{mvs,sc,jbm,hez}.  Characterization of leadership statistics is
sensible only for finite systems.  We thus consider large yet finite
systems for which the rate equation approach yields the leading
asymptotic dependence on the system size \cite{aal,jls,ve}.

Our main result is that the total number of lead changes $L$ grows as
$L(N)\sim[\ln N]^2$ with the system size $N$. This as well as other
leadership statistics are universal as they characterize both random
trees and random graphs.  The time dependent number of lead changes
$L(t,N)$ attains the scaling form $(\ln N)^2\,F(x)$ with the scaling
variable $x=\ln k_*/\ln N$ where $k_*$ is the typical component
size. Additionally, the probability that no lead change ever occurs
decays as $e^{-L}$.

We start with the simpler case of random trees. These are generated
according to the following procedure.  Initially, the system consists
of $N$ single-leaf trees.  Then, two trees are picked at random and
attached to a common root.  This merging process is repeated until a
single tree with $N$ leafs emerges. We treat the merging process
dynamically. Let $n$ be the number of trees. In the thermodynamic
limit, the normalized density $c=n/N$ evolves according to
$\frac{d}{dt}c=-c^2$ because in every merger two trees are lost and
one is gained (for convenience, the merger rate is set to
unity). Subject to the initial condition $c(0)=1$ the density is
$c(t)=(1+t)^{-1}$ and given the simple relations between the number of
trees $n=N(1+t)^{-1}$, the average size $m=1+t$, and time $t$, we
state our results in terms of time.

The size distribution is obtained similarly. Let $n_k$ be the number
of trees with $k$ leafs. The normalized density $c_k=n_k/N$ evolves
according to the Smoluchowsky rate equation
\hbox{$\frac{d}{dt}c_k=\sum_{i+j=k}c_ic_j-2cc_k$} with the
monodisperse initial conditions $c_k(0)=\delta_{k,1}$. The rate
equation reflects the fact that trees are merged randomly, independent
of their size. It can be solved (using the generating functions
technique for example) to give \cite{mvs,sc}
\begin{equation}
\label{ckt-sol}
c_k(t)=\frac{t^{k-1}}{(1+t)^{k+1}}.
\end{equation}
In the long time limit, the size distribution attains a simple
self-similar behavior
\begin{equation}
\label{ckt-scl}
c_k(t)\simeq k_*^{-2}\Phi(k/k_*),\qquad \Phi(x)=e^{-x}
\end{equation}
with the typical size $k_*\simeq t$.

{\it What is the average size of the largest tree (the leader)?} Using
the size distribution, we can answer an even more general
question. Let $l_r(t)$ be the average size of the $r$-largest tree
with the leader $l\equiv l_1$.  From the cumulative distribution
$u_k=\sum_{j\geq k} n_j=Nt^{-1}[t/(1+t)]^k$ and the relation
$u_{l_r}=r$, the size of the $r$th leader is
\begin{equation}
\label{lt-sol}
l_r(t,N)=\frac{\ln[N/rt]}{\ln[(1+t)/t]}\,.
\end{equation}

There are two regimes of behavior. In the short time limit, $t\ll 1$,
one has $l_r(t,N)=1+\ln[N/r]/\ln[1/t]$. Moreover, from $n_k\simeq
Nt^{k-1}$ the first dimer appears at $t_2\simeq N^{-1}$; the first
trimer appears at $t_3\simeq N^{-1/2}$ and then, there are of the
order $N^{1/2}$ dimers, so this trimer results from the leading dimer
with probability of the order $N^{-1/2}$. Since almost every lead
change introduces a new leader, the leader grows in increments of
unity. At the crossover point, $t\approx 1$, the size of the leader
varies logarithmically with the system size, \hbox{$l(t\approx
1,N)\sim \ln N$}. In the long time limit, $t\gg 1$, the size of the
leader grows linearly (up to a logarithmic correction) with time
\begin{equation}
\label{lt-late}
l_r(t,N)\simeq t\, \ln\frac{N}{rt}\,.
\end{equation}

\begin{figure}[t]
\centerline{\epsfxsize=7.6cm\epsfbox{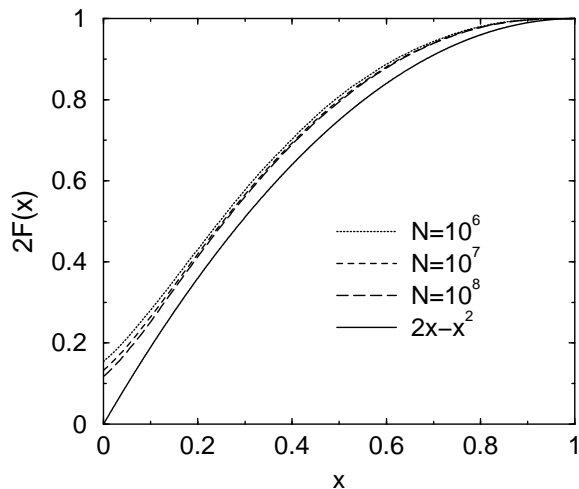}} \caption{The
normalized time dependence of the number of lead changes for random
trees, $L(t,N)/L(N)$, versus the scaling variable $x=\ln t/\ln N$.
The simulation data, representing an average over $10^3$ Monte Carlo
runs, is compared with the theoretical prediction $2F(x)=2x-x^2$.}
\end{figure}

{\it What is the average size $l_w$ of the winner (the last emerging
leader)?  At what time $t_w$ does the winner emerge?}  Both quantities
grow linearly with $N$: $l_w\simeq\alpha N$ and $t_w\simeq\beta
N$. The curve $\alpha=-\beta\ln \beta$ obtained from
Eq.~(\ref{lt-late}) has an extremum at
$\alpha=\beta=e^{-1}\cong0.36788$ thereby implying the bounds:
$\alpha,\beta<e^{-1}$.

{\it How many lead changes $L(t,N)$ occur as a function of time? As a
function of system size? What is the total number of lead changes
$L(N)\equiv L(t=\infty,N)$?}  In our definition, a lead change occurs
when two trees (none of which is the leader) merge to become larger
than the leader. For short times, $t\ll 1$, we noted that
$L(t,N)=l(t,N)-1$.  For long times, $t\gg 1$, consider the cumulative
distribution $u_k\simeq Nt^{-1}\exp(-k/t)$.  Its growth rate
immediately gives the rate by which the leader is surpassed,
$\frac{d}{dt}L(t,N)=\frac{d}{dt}u_k\big|_{k=l}$. As $u_l\simeq 1$, we
have $\frac{d}{dt}L(t,N)\simeq lt^{-2}\simeq t^{-1}\ln \frac{N}{t}$
from which the time-dependent number of lead changes is
\begin{equation}
\label{ltn-sol}
L(t,N)\simeq\ln t\ln N-\frac{1}{2}(\ln t)^2.
\end{equation}
Interestingly, this quantity obeys the scaling form
\begin{equation}
\label{ltn-scl}
L(t,N)=(\ln N)^2\,F(x), \qquad x=\frac{\ln t}{\ln N}
\end{equation}
with the quadratic scaling function: $F(x)=x-\frac{1}{2}x^2$.  The
scaling variable is unusual: a ratio of logarithms, in contrast with
the ordinary ratio underlying the size distribution (\ref{ckt-scl}).

To check these theoretical predictions, we performed large-scale Monte
Carlo simulations.  In the simulations, randomly chosen trees are
merged repeatedly. Keeping track of the leader and averaging over many
independent realizations, we observe a scaling behavior that is
consistent with Eq.~(\ref{ltn-scl}). However, as a function of the
system size, the convergence is slow because the scaling variable
involves logarithms.

\begin{figure}[t]
\centerline{\epsfxsize=7.6cm\epsfbox{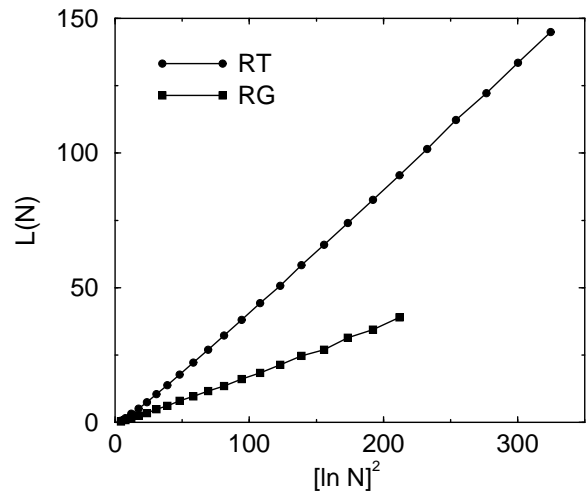}} \caption{The total
    number of lead changes $L(N)$ versus the system size $N$.  Shown
    are simulation results for Random Trees (RT) and Random Graphs
    (RG) representing an average over $10^4$ realizations. }
\end{figure}

We briefly mention a neat alternative derivation of the time-dependent
behavior.  The number of lead changes can be obtained from
$\frac{d}{dt}L=(\Delta t)^{-1}$, where the time interval between
successive lead changes $\Delta t$ is estimated from
$l_1(t)=l_2(t+\Delta t)$. This approach confirms the scaling form
(\ref{ltn-scl}) with $F(x)=(x-\frac{1}{2}x^2)/\ln 2$, i.e., there is a
factor $1/\ln 2$ discrepancy.

The time dependent behavior can be used to obtain the total number of
lead changes. Substituting $t_w=\beta N$ into Eq.~(\ref{ltn-sol})
gives
\begin{equation}
\label{ln}
L(N)\simeq A(\ln N)^2\,
\end{equation}
with $A=F(1)=1/2$. Both the leading asymptotic behavior and the $\ln
N$ correction are confirmed numerically. Moreover, the numerical
prefactor $A=0.50(1)$ agrees with the theoretical prediction
(Fig.~2). Since $L(t\approx 1,N)\sim \ln N$, the majority of lead
changes occur when $t\gg 1$.

{\it How is the number of lead changes distributed?} {\it What is the
  probability that no lead change occur?} Let $P_n(t,N)$ be the
  probability that $n$ lead changes occur till time $t$. The flux
  surpassing the leader characterizes the evolution of the probability
  distribution
  \hbox{$\frac{d}{dt}P_n=(\frac{d}{dt}L)\,[P_{n-1}-P_n]$}.  With the
  initial condition $P_n(0,N)=\delta_{n,0}$, the distribution is
  Poissonian
\begin{equation}
\label{pn}
P_n(t,N)=\frac{[L(t,N)]^n}{n!}\,e^{-L(t,N)}.
\end{equation}
Consequently, the distribution of the total number of lead changes
\hbox{$P_n(N)\equiv P_n(t=\infty,N)$} is also Poissonian:
\hbox{$P_n(N)=\frac{L^n}{n!}\,e^{-L}$} with $L\equiv L(N)$ given by
Eq.~(\ref{ln}).  Hence, the variance in the number of lead changes
$\sigma(N)$ grows as $\sigma(N)\simeq \sqrt{A}\ln N$.  Furthermore,
the probability that no lead change occur (the survival probability
of the first leader) $S(N)\equiv P_0(N)$ decays faster than a
power-law but slower than a stretched exponential
\begin{equation}
\label{sn}
S(N)=\exp[-L]\simeq \exp\left[-A(\ln N)^2\right].
\end{equation}
The asymptotic $N$-dependence is confirmed numerically (Fig.~3).

\begin{figure}[t]
\centerline{\epsfxsize=7.6cm\epsfbox{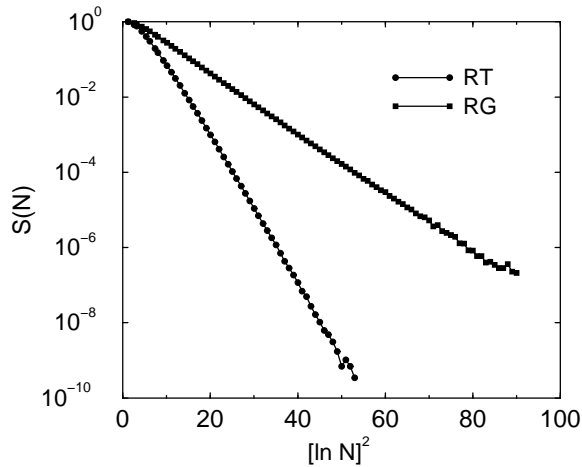}} \caption{The survival
probability of the initial leader $S(N)$ versus the system size
$N$. The number of realizations was $10^{10}$ and $10^8$ for random
trees and random graphs, respectively.}
\label{sn-fig}
\end{figure}

We now consider graphs, grown randomly as follows.  Initially, the
system consists of $N$ single-node graphs.  Then, two nodes are picked
at random and a link is drawn between them. If they belong to two
distinct graphs, the two become one. This process is repeated
indefinitely.  Let $n_k$ be the number of graphs of size $k$.  The
normalized density $c_k=n_k/N$ evolves according to the rate equation
$\frac{d}{dt}c_k=\frac{1}{2}\sum_{i+j=k}ijc_ic_j-k\,c_k$ with the
monodisperse initial conditions $c_k(0)=\delta_{k,1}$. The rate
equation reflects the fact that two graphs are connected with rate
proportional to the product of their sizes.  This equation can be
solved using generating functions to give \cite{jbm,hez}
\begin{equation}
\label{rg-ckt-sol}
c_k(t)=\frac{(kt)^{k-1}}{k\cdot k!}e^{-kt}.
\end{equation}

At time $t=1$, the system undergoes a gelation transition: it develops
a giant component that eventually takes over the entire mass in the
system.  Close to this gelation time the size distributions attains
the scaling behavior
\begin{equation}
\label{rg-ckt-scl}
c_k(t)\simeq k_*^{-5/2}\Phi(k/k_*),\qquad \Phi(x)\propto x^{-5/2}e^{-x/2}.
\end{equation}
The typical size diverges, \hbox{$k_*\simeq (1-t)^{-2}$}, as $t\to
1$. At the gelation time, the size distribution has an algebraic tail
\hbox{$c_k(t=1)\sim k^{-5/2}$}.  Hence, the cumulative distribution is
\hbox{$u_k\sim Nk^{-3/2}$} and the criterion $u_{l_w}\sim 1$ gives the
average size of the giant component (the winner), $l_w\sim N^{2/3}$
\cite{bb}.  The time at which it emerges is $1-t_w\sim N^{-1/3}$.

We estimate the size of the leader from the cumulative distribution,
$u_l=1$.  For $t\ll 1$, the size of the leader $l(t,N)$, the number of
lead changes $L(t,N)$, as well as the number of distinct leaders are
all approximately equal and the same as for random trees.  The number
of lead changes is of the order $\ln N$ in this phase; furthermore,
$L(t,N)\sim \ln N$ for $t<1$. The behavior near the gelation time
$1-t\ll 1$ is a bit more interesting. {}From the large-$k$ behavior,
\hbox{$u_k\sim N(1-t)^{-2}k^{-5/2}\exp[-k(1-t)^2/2]$}, and $u_l=1$ we
arrive at following implicit relation for the leader $l(t,N)$:
\hbox{$l\simeq 2(1-t)^{-2}\,\ln N-3(1-t)^{-2}\,\ln l\,$}.  Inserting
the zeroth order approximation \hbox{$l^{(0)}=2(1-t)^{-2}\,\ln N$}
into $\ln l$ on the right-hand side of the above relation and ignoring
$\ln\ln N$ terms yields the leader size
\begin{equation}
\label{rg-lt-late}
l\simeq\frac{2}{(1-t)^2}\,\ln [N(1-t)^3].
\end{equation}
The rate by which the leader changes is estimated from
$\frac{d}{dt}L=\frac{d}{dt}u_k\big|_{k=l}\simeq
l(1-t)$. Substituting Eq.~(\ref{rg-lt-late}) and integrating, the
number of lead changes is
\begin{equation}
\label{rg-lnt}
L(t,N)\simeq 2\,\ln N\,\ln\frac{1}{1-t}
-3\left[\ln \frac{1}{1-t}\right]^2\,.
\end{equation}
It attains the scaling form
\begin{equation}
\label{rg-ltn-scl}
L(t,N)\simeq (\ln N)^2\,F(x), \qquad
x=\frac{\ln\frac{1}{1-t}}{\ln N}
\end{equation}
with the scaling function $F(x)=2x-3x^2$.  As this behavior holds up
to time $t_w$, where $1-t_w\sim N^{-1/3}$, the total number of lead
changes grows according to Eq.~(\ref{ln}) with
$A=F(1/3)=1/3$. Furthermore, the distribution of lead changes is
Poissonian as in (\ref{pn}) and the survival probability decays
according to (\ref{sn}).

As terms of the order $\ln\ln N/\ln N$ were neglected, the scaling
behavior and the leading asymptotic behavior may be realized only for
extremely large systems. Moreover, the computational cost of random
graph simulations is larger because graphs are chosen with probability
proportional to their size. Nevertheless, we can confirm the predicted
system size dependence of $L(N)$ (Fig.~2) and $S(N)$ (Fig.~3)
numerically. The prefactor $A=0.20(2)$ is lower than the theoretical
value $A=1/3$, perhaps due to the slow convergence.

Let us compare random trees and random graphs. They seem very
different, e.g., the gelation transition occurs in one case but not in
the other. Yet, they exhibit similar extremal characteristics.  In
both cases, the total number of lead changes $L(N)$ grows as $[\ln
N]^2$ and the survival probability decays as $\exp[-L]$.  Moreover,
even the seemingly distinct temporal characteristics can be
reconciled, e.g., in both cases the size distribution attains the
scaling form $c_k(t)\propto \Phi(k/k_*)$ when $t\to\infty$ (for random
trees) and $t\to 1$ (for random graphs). Of course, the actual time
dependence of the typical scale is different: $k_*\sim t$ and $k_*\sim
(1-t)^{-2}$, respectively. The size of the leader can be rewritten as
$l\approx k_*\ln [N/k_*^\gamma]$ with $\gamma=1$ and $3/2$,
respectively.  Furthermore, Eqs.~(\ref{ltn-scl}) and
(\ref{rg-ltn-scl}) can be reconciled by writing the scaling variable
in the unified form $x=\ln k_*/\ln N$.

We now restrict our attention to the survival probability $S(N)$ and
supplement the leading behavior (\ref{sn}) with rigorous bounds.
Consider random trees. A lower bound for $S(N)$ is obtained from a
greedy scenario in which all merger events involve the leader till it
reaches size $N/2$. The probability that the second merger involves
the leading dimer is $(N-1)^{-1}$; the probability that the third
merger involves the leading trimer is $(N-2)^{-1}$; etc.  Thus, the
greedy scenario is realized with probability
$\prod_{j<N/2}\frac{1}{N-j}$, thereby providing the lower bound
$S(N)>\frac{(N/2)!}{(N-1)!}$.  An upper bound can be obtained by
estimating the number of trees the size of the leader. There are of
the order $N^{1/2}$ dimers when the first trimer is born,
$n_2(t_3=N^{-1/2})=N^{1/2}$. The leading dimer retains the lead with
probability inversely proportional to the number of dimers,
$N^{-1/2}$. Similarly, this leading trimer retains the lead with
probability proportional to $N^{-1/3}$. Therefore the upper bound
$\prod_{j<\ln N} N^{-1/j}$ is estimated as $N^{-\ln \ln N}$ (the
cutoff $j<\ln N$ is dictated by the size of the leader at the
crossover time $t\approx 1$).  Hence, the survival probability obeys
\begin{equation}
\label{bounds-rt} \left(\frac{e}{2N}\right)^{\frac{N}{2}}<S(N) <
\exp\left[-(\ln N)\cdot(\ln\ln N)\right],
\end{equation}
where the lower bound was simplified using the Stirling formula.  Note
that the upper bound merely assures that the lead never changes in the
early phase $t<1$ when the average number of lead changes is only $\ln
N$.

For random graphs, the greedy scenario is again simple to analyze
since the probability that in a system with a leader of size $j$ and
$N-j$ monomers the probability that the next merger involves the
leader is \hbox{$p_j=[j(N-j)]/[j(N-j)+\frac{1}{2}\,(N-j)(N-j-1)]$} or
$p_j=(2j)/(N+j-1)$. The product $\prod_{j<N/2}p_j$ provides the lower
bound.  Asymptotically, the lower bound decays as $\lambda^N$ with
$\lambda=(2/3)^{3/2}=0.544331\ldots$. On the other hand, repeating the
above argument yields the same upper bound, so
\begin{equation}
\label{bounds-rg}
\lambda^N<S(N)<\exp\left[-(\ln N)\cdot(\ln\ln N)\right]\,.
\end{equation}
The upper bound is again much closer to the actual asymptotic behavior.

In conclusion, random graphs and random trees exhibit similar
leadership characteristics. As in random growing networks \cite{kr},
lead changes are infrequent given that the overall number of lead
changes increases only logarithmically with the system size.  The time
dependent number of lead changes approaches a self-similar form
asymptotically. The convergence to the asymptotic behavior is much
slower for extremal statistics compared with size statistics due to
the various logarithmic dependences.  Consequently, the asymptotic
behavior may difficult to detect in practice, especially for random
graphs.

To obtain the extremal characteristics, we employed the scaling
behavior of the size distribution outside the scaling regime, namely,
at sizes much larger than the typical size where, at least formally,
statistical fluctuations can no longer be ignored.  Interestingly, the
resulting system size dependence for the various leadership statistics
appears to be asymptotically exact. Further analysis is needed to
illuminate the role of statistical fluctuations, for example by
characterizing corrections to the leading behavior.

The virtue of the rate equation approach to analyzing extremal
characteristics is its simplicity, robustness, and generality. It
applies to general aggregation processes where the merger rate may
depend in a complicated manner on the aggregate size or in situations
where there is an underlying spatial structure. We find that the above
leadership statistics extend to algebraic merger rates as well as to
aggregation in one spatial dimension. This method is also applicable
to other extremal features including for example laggard (smallest
component) statistics. In the case of random trees, for instance, the
total number of laggard changes grows logarithmically with the system
size.

\acknowledgments We are thankful to Sidney Redner for many discussions
and initial collaboration on this work. This research was supported in
part by DOE(W-7405-ENG-36).

\end{document}